\documentclass[a4paper,12pt]{article}

\usepackage[dvips]{graphicx}
\usepackage{pstricks}
\usepackage{bm}

\usepackage{graphicx}

\newcommand{\be}{\begin{equation}}
\newcommand{\ee}{\end{equation}}

\newcommand{\ka}{\kappa}

\newcommand{\beq}[1] {\begin{equation}\label{#1} }
\newcommand{\eeq} {\end{equation} }
\newcommand{\bea}[1]{\begin{eqnarray}\label{#1} }
\newcommand{\eea}{\end{eqnarray}}

\newcommand{\si}{\sigma}

\def\beqn{\begin{eqnarray}}
\def\eeqn{\end{eqnarray}}

\def\beq{\begin{equation}}
\def\eeq{\end{equation}}
\def\bea{\begin{equation}}
\def\eea{\end{equation}}

\def\al{\alpha}
\def\bt{\beta}
\def\Ga{\Gamma}
\def\ga{\gamma}
\def\de{\delta}
\def\De{\Delta}

\def\ka{\kappa}
\def\si{\sigma}
\def\Si{\Sigma}
\def\te{\theta}

\def\lam{\lambda}

\def\om{\omega}
\def\ep{\epsilon}

\def\sq{\sqrt}

\def\l{\left (}
\def\r{\right )}

\def\fr{\frac}
\def\la{\label}
\def\hs{\hspace}
\def\vs{\vspace}

\def\ran{\rangle}
\def\lan{\langle}
\def\ov{\overline}
\def\tl{\tilde}
\def\tm{\times}

\def\orvec{\overrightarrow}

\begin{document}

\begin{flushright}
OSU-HEP-10-01\\
SLAC-PUB-13732\\
March 12, 2010 \\
\end{flushright}

\begin{center}
{\Large\bf Constraining Proton Lifetime in $SO(10)$ with \\[0.1in]
 Stabilized Doublet-Triplet Splitting}\\
\end{center}

\vspace{0.5cm}
\begin{center}
\renewcommand{\thefootnote}{\fnsymbol{footnote}}
{\large
{}~K.S. Babu$^{\hs{0.5mm}a}$\footnote{E-mail: babu@okstate.edu},{}~
Jogesh C. Pati$^{\hs{0.5mm}b}$\footnote{E-mail: pati@slac.stanford.edu}, and
{}~Zurab Tavartkiladze$^{\hs{0.5mm}a, c}$\footnote{E-mail: zurab.tavartkiladze@gmail.com}
}
\vspace{0.5cm}

$^a${\em Department of Physics, Oklahoma State University, Stillwater, OK 74078, USA }

$^b${\em SLAC National Accelerator Laboratory, 2575 Sand Hill Road, Menlo Park, CA 94025, USA }

$^c${\em E. Andronikashvili Institute of Physics, Tamarashvili 6, Tbilisi 0177, Georgia}
\end{center}

\begin{abstract}

We present a class of realistic unified models based on supersymmetric $SO(10)$ wherein
issues related to natural doublet-triplet (DT) splitting are fully resolved.  Using a minimal
set of low dimensional Higgs fields which includes a single adjoint, we show that the
Dimopoulos--Wilzcek  mechanism for DT splitting can be made stable in the presence of all higher order
operators without having pseudo-Goldstone bosons and flat directions.
The $\mu$ term of order TeV is found to be naturally induced. A $Z_2$-assisted anomalous
${\cal U}(1)_A$ gauge symmetry plays a crucial role in achieving these results.
The threshold corrections to $\al_3(M_Z)$, somewhat surprisingly, are found to be controlled by only a few effective parameters. This leads to a very  predictive scenario for proton decay. As a novel feature, we find  an interesting correlation between the $d=6$ ($p\to e^+\pi^0$) and $d=5$ ($p\to \ov{\nu }K^+$) decay amplitudes which allows us to derive a constrained upper limit on the inverse rate of the $e^+\pi^0$ mode. Our results show that both modes should be observed with an improvement in the current sensitivity by about a factor of five to ten.

\end{abstract}

\newpage

\section{Introduction}\label{sec:intro}

Although yet to be seen, proton decay is an indispensable tool to probe nature at truly high energies ($\sim 10^{16}$~GeV).
It still remains as the missing piece of grand unification \cite{Pati:1974yy, GG-gut, Georgi:1974yf}. In the light
of a new set of planned detectors including those at the forthcoming
deep underground laboratory DUSEL \cite{Raby:2008pd} and the HyperKamiokande, we propose to address
in this paper certain well known but partially unresolved theoretical issues of supersymmetric (SUSY) grand unification (GUT)
which are especially relevant to proton decay.

Strong empirical support for grand unification arises not only from the
observed quantum numbers of quarks and leptons and the quantization of electric charge, but in particular from the meeting of the three gauge couplings  at a scale $M_{\rm GUT} \approx 2\cdot 10^{16}$~GeV
that occurs in the context of low energy SUSY \cite{Dimopoulos:1981yj},
and the tiny neutrino masses, as observed in neutrino oscillation experiments. The latter fit extremely well with
GUT symmetries that include the symmetry $SU(2)_L\tm SU(2)_R\tm SU(4)_C$ \cite{Pati:1974yy}, the minimal such
symmetry being $SO(10)$ \cite{Fritzsch:1974nn}.
 We will therefore discuss  proton decay in the context of supersymmetric $SO(10)$.
The purpose of the present paper is to pay special
attention to the problem of the so-called doublet-triplet (DT) splitting and to study the implications of its resolution  for proton decay.

The DT splitting problem is common to all  grand unified theories based on  simple gauge groups.
In SUSY $SO(10)$ models the two Higgs doublets of MSSM, a color triplet and an anti-triplet
lie (typically) in a $10$-dimensional  representation $H(10)$. The  color triplets need
to be superheavy  so as to avoid rapid proton decay and  also to preserve gauge coupling unification. Keeping
the doublets light and the triplets superheavy  self-consistently is the  doublet-triplet  splitting problem.

A natural solution to this problem, avoiding severe fine-tuning is realized in
SUSY $SO(10)$  by the so called Dimopoulos-Wilczek (or the missing VEV) mechanism \cite{Dimopoulos:1981xm}.
It involves a coupling of two $10$-plets of the form $H(10)A(45)H'(10)$ with the adjoint  $A(45)$ having a GUT scale VEV in
the $(B-L)$-preserving direction:
\beq
\lan A\ran ={\rm i}\si_2\otimes {\rm Diag}\l a, ~a,~a,~0,~0\r ~.
\la{AVEV-BL}
\eeq
This structure contributes to the
triplet and not to the doublet masses, and thereby can lead to natural DT splitting without fine-tuning.

Given the very large hierarchy between the doublet and triplet masses, however,
one must ensure: (i) that the missing VEV pattern for $A(45)$ in Eq. (\ref{AVEV-BL}) is stable to a high enough accuracy
in the presence of all allowed higher dimensional operators;
(ii) that there are no undesirable pseudo-Goldstone bosons; and
(iii) that there are no flat directions which would lead to VEVs of fields undetermined.
Furthermore, (iv) one must also examine, by including all GUT-scale threshold corrections to the gauge couplings,
 the implication of the doublet-triplet splitting on coupling unification and on proton decay.
 To our knowledge, while some of these issues have been partially addressed in the literature
 (e.g. see \cite{more-mis-VEV}, \cite{Babu:1998wi},  \cite{Berezh}), and major
 progress was made in Ref. \cite{Barr:1997hq} with regard to the issues (i) and (ii), {\it simultaneous resolution} of all four issues
 has so far remained a challenge.

In this paper we present a predictive class of $SO(10)$ models, based on a minimal Higgs
system, in which all the issues of DT splitting mentioned above are resolved, and where the threshold corrections to the gauge couplings and
their implications for proton decay are properly studied as well.  The Higgs sector we consider has a single adjoint, along with vectors and spinors.  Such a low dimensional Higgs system would lead to smaller threshold effects [8,10], unlike in  models  \cite{more-mis-VEV,Berezh} which employ multiple adjoints and/or 54 dimensional Higgs.\footnote{An alternative class of $SO(10)$ models utilizing larger dimensional
(e.g. 126) Higgs fields has been studied in Ref. \cite{n-minSO101}.  These models have
the interesting feature that $R$--parity is automatic, being part of the gauge symmetry.  However, threshold
corrections are rather large in these models, making quantitative predictions
for $\alpha_3(m_Z)$ and proton decay difficult (see  attempts in this regard by Aulukh and Garg \cite{n-minSO101}).}
A postulated $Z_2$-assisted  anomalous ${\cal U}(1)_A$ symmetry (which may have a string origin \cite{Dine:1987xk}) plays a crucial role
in obtaining our results.
We find somewhat surprisingly that the GUT scale threshold corrections to $\al_3(M_Z)$ are determined in terms of a very few parameters.
This makes the model rather predictive for proton decay.
As a novel feature, we find an intriguing {\it correlation} between the $d=6$ and $d=5$
proton decays, which respectively lead to $p\to e^+\pi^0$ and $p\to \ov{\nu }K^+$ as the dominant decay modes. The correlation is such that the
empirical lower limit on
$\Ga^{-1}(p\to \ov{\nu }K^+)$ provides {\it a constrained upper limit} on $\Ga^{-1}(p\to e^+\pi^0)$.
Our results show that both decay modes should in fact be discovered
with an improvement in the current limits on lifetimes by about a factor of five to ten.

\section{Stabilizing doublet-triplet splitting}

In order to break SUSY $SO(10)$ to the supersymmetric standard model with a stabilized DT sector, and for the subsequent breaking of the electro-weak symmetry, we shall use a minimal low dimensional Higgs system. It consists of
a single adjoint $A(45)$, two pairs of spinor-antispinor superfields $\{C(16)+ \bar C(\ov{16})\}$ and $\{C'(16)+\bar C'(\ov{16})\}$,  two $10$-plets $H(10)$ and $H'(10)$, as well as two $SO(10)$ singlets $S$ and $Z$.
The second spinorial pair $C'+\bar C'$ is introduced, following Ref. \cite{Barr:1997hq},
to avoid pseudo-Goldstone degrees of freedom while maintaining the  Dimopoulos-Wilczek VEV structure for $A$ (cf: Eq. (\ref{AVEV-BL})).
The $S$ and $Z$ superfields are needed to fix various VEVs  in the required directions through their superpotential couplings.

We supplement the gauge symmetry by a $Z_2$-assisted  anomalous ${\cal U}(1)_A$  symmetry in order to stabilize the VEV pattern of Eq. (\ref{AVEV-BL})
 \cite{pgb-anU1}, \cite{Berezh}.
The charges of the Higgs fields and those of the three matter families $16_i$ under ${\cal U}(1)_A\tm Z_2$ are listed in  Table \ref{t:U-Z-charges}.
%
%
%
%
\begin{table}[h]
\caption{${\cal U}(1)_A$ and $Z_2$ charges $Q_i$ and $\om_i$ of the superfield $\phi_i$. The transformations under
${\cal U}(1)_A$ and $Z_2$ are respectively $\phi_i\to e^{iQ_i}\phi_i$ and $\phi_i\to e^{i\fr{2\pi }{2}\om_i}\phi_i$.}
\vs{-0.5cm}
\label{t:U-Z-charges} $$\begin{array}{|c||c|c|c|c|c|c|c|c|c|c|c|}
\hline
\vs{-0.7cm}
 &  &  &  &  &  &  &  & & &&\\

\vs{-0.7cm}

& ~A(45)~& ~H(10)~&~ H'(10)~& ~C(16) ~& ~ \bar C(\ov{16})~&~Z ~&~ S ~  & ~C'(16) ~ & ~\bar C'(\ov{16}) & 16_{1,2}&16_3\\

&  &  &  &  &  &  &  &  & &&\\

\hline

\vs{-0.7cm}
 &  &  &  &  &  &  &  & & &&\\

~Q~& 0 &1  &-1  &\fr{k+4}{2k}  & -\fr{1}{2}   &\fr{2}{k}   & \fr{2}{k}   & \fr{k-4}{2k}  & -\fr{k+8}{2k}   &q_{1,2}&-\fr{1}{2}\\

\vs{-0.7cm}
&  &  &  &  &  &  &  &  & &&\\

\hline
\vs{-0.7cm}
 &  &  &  &  &  &  &  & &&&\\
\vs{-0.7cm}
~\om ~& 1 & 0 & 1 & 0 & 0  & 1  & 0  & 0  & 0  &P_{1,2}&0\\

&  &  &  &  &  &  &  &  & &&\\

\hline

\end{array}$$

\end{table}
%
%
%
%
%
 Here $k$ is a positive integer which is unspecified for the moment. The superpotential of the symmetry breaking sector, consistent with these symmetries, is
$W=W_1+W_2+W_3$, where
\begin{eqnarray}
W_1\hs{-0.25cm}&=&\hs{-0.25cm}M_A{\rm tr}A^2+\fr{\lam_A}{M_*}\l {\rm tr}A^2\r^2+\fr{\lam_A'}{M_*} {\rm tr}A^4~,
\la{sup-A}\\
W_2\hs{-0.25cm}&=&\hs{-0.25cm}C\hs{-0.8mm}\l \fr{a_1}{M_*}ZA+\fr{b_1}{M_*}C\bar C+
c_1S\r \hs{-0.8mm}\bar C' +C'\hs{-0.8mm}\l \fr{a_2}{M_*}ZA+\fr{b_2}{M_*}C\bar C+c_2S\r \hs{-0.8mm}\bar C ,
\la{sup-ACCbar}\\
W_3\hs{-0.2cm}&=&\hs{-0.2cm}\lam_1HAH'\hs{-0.1cm}+\hs{-0.1cm}\hs{-0cm}\l \lam_{H'}SZ^{k-1}+ {\lam'}_{H'}Z^k\r \hs{-0.1cm}\fr{(H')^2}{M_*^{k-1}}
\hs{-0cm}+\lam_2H\bar C\bar C \hs{-0.1cm}+\hs{-0.1cm}\fr{\lam_3}{M_*}AH'CC'~.
\la{W-H}
\end{eqnarray}
For simplicity we assume that the $SO(10)$ contractions in the $C\bar C$ terms with coefficients  $b_{1,2}$  in Eq. (\ref{sup-ACCbar}) are in the singlet channel. In the second term of
Eq. (\ref{W-H}) the operator $Z^k$ can  appear only when $k$ is even.
While our consideration of DT splitting will hold for all $k$, if $k$ is odd, matter parity is automatic, being part of ${\cal U}(1)_A$.
The choice of $k=5$, which we will use, is phenomenologically preferred, in particular for suppressing adequately all $d=5$ proton decay operators
including those induced by Planck scale physics.
Higher order operators such as $A^6/M_*^3$ etc. are not exhibited in Eqs. (\ref{sup-A})-(\ref{W-H}) because they are inconsequential for our purposes.
The charges $q_{1,2}$ and the parities $P_{1,2}$  of the first two families are left unspecified for the present. They will however be relevant for the generation of quark and lepton masses.

Typically, we expect that the non-renormalizable operators such as $\lam_A$ and $\lam_A'$-terms would be induced by quantum gravity effects
involving exchange of heavy states in the string tower.  Thus, we expect the cut-off scale $M_*\sim M_{\rm Pl}$ or
$M_{\rm String}\sim 10^{18}$~GeV.  We shall take all dimensionless couplings to be of order unity, i.e., in the
range $(1/4-2)$.

Using the SUSY preserving condition $F_Z=F_S=0$, together with the choice
 $\left\langle C \right \rangle = \left \langle \bar{C} \right \rangle = c,
 \lan A\ran  \neq 0$ (which is one allowed option among the discrete set of degenerate vacuum solutions), we get $\lan C\bar C'\ran =\lan \bar CC'\ran =0$ and $\lan C'\ran = \lan \bar C'\ran =0$.
 The VEV of $A$ is then determined entirely by
$W_1$ of Eq. (\ref{sup-A}). Setting $F_A=0$, we find a solution  in the $B-L$ direction as in Eq. (\ref{AVEV-BL}), with
\begin{equation}\label{a2}
a^2=\fr{M_AM_*}{2(6\lam_A+\lam_A')}~.
\end{equation}
With $\lam_A, \lam_A'\sim 1$  and $M_*\sim 10^{18}$~GeV, we need to choose $M_A\sim 10^{15}$~GeV
to obtain $a\sim M_{\rm GUT}\approx 2\cdot 10^{16}$~GeV.
Demanding $F$-flatness conditions $F_{C'}=F_{\bar C'}=0$ and using
the notations $z=\lan Z\ran $ and $s=\lan S\ran $
we get $s=\fr{c^2}{M_*}\rho_1$ and $z=\fr{c^2}{3a}\rho_2$, where
$\rho_1=\fr{b_1a_2-b_2a_1}{a_1c_2-a_2c_1}~,~\rho_2=\fr{b_1c_2-b_2c_1}{a_1c_2-a_2c_1}$. We note that for all dimensionless couplings in the Lagrangian being in the range $(1/4-2)$, the effective couplings $\rho_{1,2}$ can naturally take values as small as about $1/50$.

The sum of the VEVs gets further constrained as follows.
The anomalous ${\cal U}(1)_A$ symmetry, presumed to have a string origin, generates the Fayet-Iliopoulos
term $\xi $ through quantum gravity, which is given by \cite{Dine:1987xk}
$\xi =\fr{g_{\rm st}^2M_{\rm Pl}^2}{192\pi^2}{\rm Tr}Q_A$,
where $g_{\rm st}$ denotes the string coupling and
$M_{\rm Pl}\simeq 2.4 \cdot 10^{18}$~GeV is the reduced Planck mass.
In our model, the particle spectrum of Table 1 would lead to
Tr$(Q_A) = -8 -60/k + 16(q_1 +q_2) = -84/5$, (for $k=5$, $q_{1,2} =
-1/2 + 3/k$, see later).   This value will however be modified if
there are additional singlets in the full theory.
(Semi-realistic string solutions \cite{Faraggi:1997be}, possessing an anomalous ${\cal U}(1)_A$, typically lead to $|{\rm Tr}Q_A|\approx 30-100$.)
With the charges  in Table \ref{t:U-Z-charges}, the vanishing of  $D_A=\xi +\sum_iQ_i|\lan \phi_i\ran |^2=0$ (required for preserving SUSY),
yields $c^2+|z|^2+|s|^2=-\fr{k}{2}\xi $.
Thus, the VEVs of all the fields get determined. We see that quite naturally,  the VEVs $c, z\sim ({\rm few}-10)\tm M_{\rm GUT}$, and
$s\sim (10^{-2}-10^{-1})\tm M_{\rm GUT}$ can arise, with the precise values depending on the order one couplings. Let us note that this setup also allows for additional singlet fields $\{P_i\}$ which can play a role in the $D_A=0$ condition (for $P_i$ with positive ${\cal U}(1)_A$ charges) and can modify these estimates somewhat, without upsetting the stability of DT splitting.

Substituting the VEVs of the heavy fields in Eqs. (\ref{sup-ACCbar}) and (\ref{W-H}), we  derive the mass matrices $M_D$ and $M_T$  for the $SU(2)_L$ doublets
and $SU(3)_c$-color triplets (written in the $SU(5)$ notation):
\beq
\begin{array}{cccc}
 & {\begin{array}{cccc}
 \hs{-0cm}5_H&\hs{0.8cm} 5_{H'}& \hs{0.8cm}5_{\bar C}& \hs{0.8cm} 5_{\bar C'}
\end{array}}\\ \vspace{1mm}
\hs{0.1cm}M_{D,T}= \hs{-0.3cm}
 \begin{array}{c}
\bar 5_H~\\~ \bar 5_{H'} ~ \\~ \bar 5_C ~\\ ~\bar 5_{C'}
 \end{array}\!\!\!\!\!
\hs{-0.3cm} &{\left(\begin{array}{cccc}

 0 & \eta_{D,T}\lam_1a  &\lam_2c & 0
\\
-\eta_{D,T}\lam_1a &M_{H'} &0 &0
 \\
 0&0 & 0 &\ka_{D,T}Y_1
 \\
  0&Y_{D,T} & \ka_{D,T}Y_2 &M_{C'}
\end{array}\right)}~,
\end{array}  \!\!  ~~~~~
\label{MDT}
\eeq
with $(\eta_D, \eta_T)=(0, 1)$, $(\ka_D, \ka_T)=(3, 2)$.
Here $M_{H'}\!=\!(\lam_{H'}SZ^{k-1}+ {\lam'}_{H'}Z^k)/M_*^{k-1}$
(see Eq. (\ref{W-H})), $Y_{1,2}=\!2a_{1,2}za/(M_*)$ and $Y_{D,T}\sim \lam_3\lan A\ran c/(M_*)$.
For $k=5$ the dominant contribution to $M_{H'}$  comes from the operator $\lam_{H'}sz^4/M_*^4$, which is in the range $(10^{11}-10^{12})$~GeV. The suppressed mass of $H'$ will be crucial for an adequate suppression of $d=5$ proton decay.
The entry $M_{C'}$ in Eq. (\ref{MDT}) (allowed by the stability of Higgs doublet mass)  would arise from the operator
$SZ^2C'\bar C'/M^2$ and yields $M_{C'}\!\sim \!(10^{-2} ~{\rm to}~ 10^{-1})\tm M_{\rm GUT}$  if $M\sim z$, which happens if the superfields that are integrated out have GUT scale masses.

The zeros in the first column of Eq. (\ref{MDT}) are ensured, in the presence of {\it all higher dimensional operators}, for the doublet mass
matrix by the ${\cal U}(1)_A\tm Z_2$ symmetry.
The main reason for this all-order stability of the Higgs doublet masses is that all the effective Higgs fields
(i.e. any positive power of $Z$, $S$ and $\bar CC$)  which have super-large VEVs are positively charged under
${\cal U}(1)_A$, and can not couple to $H^2$, which is also positively charged.
Thus, with $\eta_D=0$ one pair of the Higgs doublets will be massless, while the remaining three pairs of
doublets become superheavy.
The role of the $Z_2$ symmetry is that it allows the coupling of $H$ to $H'$ only through $A$ (or odd powers of $A$). Such couplings, however,
do not generate a doublet mass due to the VEV structure in Eq. (\ref{AVEV-BL}) of $\lan A\ran $.
The VEV pattern of $\lan A\ran $ along the $B-L$ direction is also guaranteed to be stable because of the ${\cal U}(1)_A$ symmetry.
Indeed, note that the symmetry ${\cal U}(1)_A$ does not allow any superpotential coupling involving $A, C$ and $\bar C$ of the form
$A^n(C\bar C)^m$. It is only these couplings which, if allowed,
would have upset the missing VEV pattern of Eq. (\ref{AVEV-BL}). Their absence to all orders thus guarantees that the pattern
of Eq. (\ref{AVEV-BL}) is absolutely stable (barring of course SUSY breaking at the TeV scale which is safe).
As far as the color-triplets are concerned, since $\eta_T \neq 0$ in Eq. (\ref{MDT}) for the triplets,
 all four pairs become super-heavy, just as desired.

The two massless Higgs doublets which emerge from Eq. (\ref{MDT}) represent the MSSM doublets $h_u$ and $h_d$
which acquire light masses after SUSY breaking. Let us denote the down type doublets in $H$, $H'$, $C$ and $C'$ by
$H_d$, ${H_d}'$, $C_d$ and ${C_d}'$ respectively, and likewise the up-type doublets. It is easy to see from Eq. (\ref{MDT}) that $h_u$
is composed entirely of $H$ - i.e. $h_u=H_u$, while $h_d$ is a mixture of four components $H_d, {H_d}', C_d$ and ${C_d}'$.
In particular, the weights of $h_d$ in $H, H', C$ and $C'$ are given by
$H\supset \cos \ga \cdot h_d$, $H'\supset \fr{\lam_2cY_D}{3Y_2M_{H'}}\cos \ga \cdot h_d$,
$C\supset \fr{\lam_2cM_{C'}}{9Y_1Y_2}\cos \ga \cdot h_d$ and $C'\supset \fr{\lam_2c}{3Y_2}\cos \ga \cdot h_d$.
The angle $\ga $ is determined in terms of the parameters of the superpotential.
It is related to the MSSM parameter $\tan \bt $ as $\tan \bt =\fr{m_t}{m_b}\cos \ga $.
Note that, unlike in many $SO(10)$ models, the MSSM parameter  $\tan\beta$ is not required to be large here.
It would turn out that conservative upper limits on proton lifetime would correspond to smaller values of $\tan\beta $.

The $\mu $-term, the coefficient of $h_uh_d$ term of MSSM superpotential,  is generated within our model in a simple way.
In the unbroken SUSY limit, $\mu $-term is zero since terms such as $H^2$
are forbidden.
After SUSY breaking, the adjoint $A(45)$ develops a VEV$\sim m_{susy}$ along its $I_{3R}$ direction, correcting the zeros of Eq. (\ref{AVEV-BL}), which generates the $\mu $-term. This occurs since the inclusion of
the soft SUSY breaking terms induces VEVs $\sim m_{susy}$ for the fields $C'$ and $\bar C'$ along their $\nu^c$-like scalar components.
These will trigger the VEV ($\sim m_{susy}$) of $A(45)$ in the $I_{3R}$ direction.
From Eq. (\ref{W-H}) we obtain $\lam_1HAH'\to m_{susy}h_uh_d$,
and thus $\mu \sim m_{\rm susy}\sim$~TeV, independent of the integer $k$.
Thus the present setup provides a simple and elegant solution to the $\mu $
problem without any new ingredients.

Using Eq. (\ref{MDT}), for the four heavy triplets $T_i$ and three heavy doublets $D_i$ (coming from four pairs of $(5+\bar 5)$'s of $SU(5)$
in $H, H', C, C', \bar C, \bar C'$) we derive the following mass relations:
\beq
\fr{M_{D_1}M_{D_2}M_{D_3}}{M_{T_1}M_{T_2}M_{T_3}M_{T_4}}=\fr{9}{4M_{\rm eff}\cos \ga }~,~~
~~~~~~\fr{1}{M_{\rm eff}}=\l M_T^{-1}\r_{11}=\fr{M_{H'}}{\lam_1^2a^2}~.
\la{DTmasRatio}
\eeq
In our model, for $k=5$ we have $M_{\rm eff}=\fr{\lam_1^2a^2}{\lam_{H'}sz^4}M_*^4$.
Putting $a^2=M_X^2/g^2$ with $g^2 \approx g_{\rm GUT}^2\approx 0.63(1\pm 0.10)$, $M_X \approx (0.6-1) \times 10^{16}$ GeV (see discussion  after Eq. (\ref{p-pi-nu})),
and taking an explicit solution\footnote{Details of estimating the VEVs based on explicit solutions to the $D_A = 0$ condition will be presented in a
forthcoming longer
paper \cite{inprep}.} for the VEVs, $z\sim 0.17 M_*$ and $s \sim M_*/70$, together with natural values of the couplings $\lambda_1 \approx (1/4-\sqrt{2})$ and
$\lambda_{H'} \approx (1/4-2)$, we estimate
\begin{equation} \label{Meff-range}
M_{\rm eff} \sim (5 \times 10^{16} - 6 \times 10^{19})~{\rm GeV}~.
\end{equation}
The mass scale $M_{\rm eff}$ will control the $d=5$ proton decay amplitude (see e.g. Ref. \cite{Babu:1998wi}).  It
would also enter the threshold corrections.  Note that $M_{\rm eff}$, which does not represent the physical
mass of any particle, can naturally exceed even $10^{19}$ GeV.

Now, the multiplets $A, C, C', \bar C$ and $\bar C'$ contain three pairs of  $(10+\ov{10})$'s of $SU(5)$, which get masses through Eqs. (\ref{sup-A})
and (\ref{sup-ACCbar}). Their mass matrix is given by:

\beq
\begin{array}{ccc}
 & {\begin{array}{ccc}
\hs{-0.6cm} \ov{\Psi }_A^{\ov{10}}& \hs{0.2cm} \ov{\Psi }_{\bar C}^{\ov{10}}& \hs{0.2cm} \ov{\Psi }_{\bar C'}^{\ov{10}}
\end{array}}\\ \vspace{1mm}
M(\Psi^{10})= \hs{-0.3cm}
\begin{array}{c}
\Psi_A^{10} \\ ~\Psi_C^{10}  \\~\Psi_{C'}^{10}
 \end{array}\!\!\!\!\!\hs{-0.1cm}&{\left(\begin{array}{ccc}

 M_{\Psi } & 0 &X_1
\\
0&0 &\ka_{\Psi }Y_1
 \\
 X_2& \ka_{\Psi }Y_2& M_{C'}
\end{array}\right)}~,
\end{array}  \!\!
\label{mass10frag}
\eeq
with $\Psi =(u^c, q, e^c)$, $\ka_{\Psi }=(2, 1, 0)$, $M_{\Psi }=(0, 0, M_{\Si }/2)$,
  where  $X_{1,2}=4a_{1,2}zc/M_*$ and $Y_{1,2}$ are defined after Eq. (\ref{MDT}).
  $M_{\Si }=2\lam_A'M_A/(6\lam_A+\lam_A')$ yields the mass of the color octet and $SU(2)_L$ triplet in $A(45)$: $M_{\Si}\equiv M_8=2M_3$.
   We see from Eq. (\ref{mass10frag}) that two pairs of $(u^c,~q,~e^c)$-like
states  are massive. The third massless pair ($10+\ov{10}$ of $SU(5)$) is eaten by the corresponding massive gauge
superfields of $SO(10)/SU(5)$.
Denoting the masses of the components of $(10+\ov{10})$'s by curly symbols (e.g. ${\cal U}^c_1\equiv M(u^c_1)$ etc.), we derive from
Eq. (\ref{mass10frag}):
${\cal U}^c_1{\cal U}^c_2=Y_1Y_2(4+p^2)$, ${\cal Q}_1{\cal Q}_2=Y_1Y_2(1+p^2)$, ${\cal E}^c_1{\cal E}^c_2=Y_1Y_2\hat{p}^2$
where $p^2=\fr{4c^2}{a^2}=\fr{|X_1|^2}{|Y_1|^2}=\fr{|X_2|^2}{|Y_2|^2}$, $\hat{p}^2=p^2\left |1-\fr{M_{\Si }M_{C'}}{2X_1X_2}  \right |$.
These expressions will be useful for the computation of threshold corrections in the model.

The masses of the heavy gauge boson superfields
corresponding to the broken generators of $SO(10)$ are given by (see e.g., the second paper in Ref. \cite{more-mis-VEV} and \cite{Babu:1998wi}):
$M^2(X,Y)=g^2a^2\equiv M_X^2$, $M^2(X',Y')=M_X^2(1+p^2)$,
$M^2(V_{u^c,\bar u^c})=M_X^2(4+p^2)$,
$M^2(V_{e^c,\bar e^c})=M_X^2p^2$
where $g$ is the unified gauge coupling at the GUT scale.
Given the $p$ and $\hat{p}$ dependence of the masses given above, we see that, except for the $e^c$-like states, threshold corrections from all other states in the $(10+\ov{10})$ matter sector cancel precisely against those from the corresponding states in the gauge sector.
 An accidental $N=4$ supersymmetry present in the model (the gauge bosons and three pairs of matter fields in the $(10+\ov{10})$ sector form an $N=4$ SUSY gauge multiplet) is responsible for this cancelation.
This results in an enormous reduction of the parameters, rendering the model very predictive for proton decay.

We have presented the whole spectrum of the theory, except for the
singlet sector, which is not relevant for the calculation of threshold
corrections.  We have however, analyzed the singlet sector and verified that
there are no unwanted pseudo-Goldstone states in the model.  While
it might appear that there is a $U(1)$ symmetry associated with the
``integer" $k$ in Table 1, it turns out that this is a linear combination
of ${\cal U}(1)_A$ and $B-L$, and its breaking does not lead to
a pseudo-Goldstone boson.

\begin{figure}
\begin{center}
\leavevmode
\leavevmode
\vspace{-0.5cm}
\includegraphics{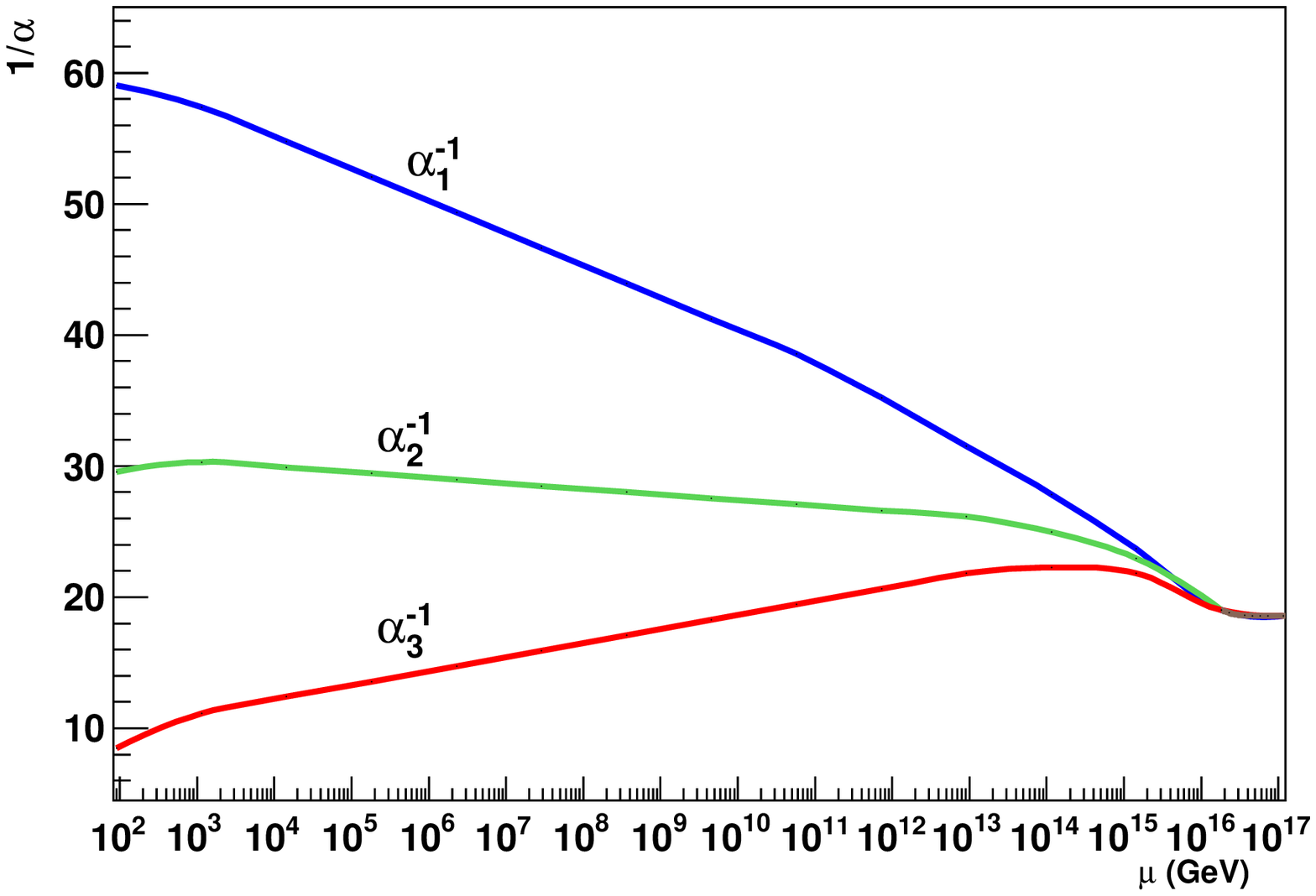}  
\end{center}
\vs{8.45cm}
\caption{Evolution of the three standard model gauge couplings in the present $SO(10)$ model including threshold
corrections.  We have used $\al_3(M_Z)= 0.1176$ and assumed an mSUGRA spectrum with $\{\tan \bt , m_0, m_{1/2}, \mu  \}=\{3, 1448.2~{\rm GeV}, 155.93~{\rm GeV}, 1~{\rm TeV} \}$
(corresponding to $m_{\tilde{q}} = 1.5$ TeV, $m_{\tilde{W}} = 130$ GeV), and have taken $p=4$, $r=1/250$, $M_{\rm eff}=4\cdot 10^{19}$~GeV, $Y_{1,2}=2M_X/45$ for generating this plot.}
\label{f:unif}
\end{figure}

The evolution of the three gauge couplings in the model with momentum is shown in
Fig. 1, which takes into account all the threshold effects.  It is clear from Fig. \ref{f:unif}
that the three couplings merge into one at a unification scale $M_{\rm GUT} \sim
10^{16}$ GeV.  Furthermore, we see that the unified $SO(10)$ gauge coupling remains
perturbative to scales well above $M_{\rm GUT}$.  This is a desirable feature which
not all $SO(10)$ models have.

\vs{-0.4cm}
\section{Novel correlation between $d=5$ and $d=6$ proton decays}
\vs{-0.3cm}

Writing down the three RG equations for $\al_{1,2,3}^{-1}$ and eliminating  the unified gauge coupling $\al_G$ we obtain
\beq
\ln \fr{M_{\rm eff}\cos \ga }{M_Z}\hs{-0.1cm}=\hs{-0.1cm}\fr{5\pi }{6}\hs{-0.1cm}\l \hs{-0.1cm}3(\al^{-1}_2+\De_{2,\hs{0.1cm}w}^{(2)}
\hs{-0.1cm}-\fr{1}{6\pi })
\!-\!2( \al^{-1}_3\!+\!\De_{3,\hs{0.1cm}w}^{(2)}\!-\!\fr{1}{4\pi })\!-\!(\al^{-1}_1+\De_{1,\hs{0.1cm}w}^{(2)})\!\r
\hs{-0.1cm}-\ln \fr{4\ka^{5/2}}{9}+\ln \fr{p}{\hat{p}}~,
\la{Meff}
\eeq
\beq
\ln \fr{\l M_X^2M_{\Si }\r^{1/3}}{M_Z}\hs{-0.1cm}=\hs{-0.1cm}\fr{\pi }{18}\hs{-0.1cm}\l \!5(\al^{-1}_1\!+\!\De_{1,\hs{0.1cm}w}^{(2)})
\!-\!3( \al^{-1}_2\!+\!\De_{2,\hs{0.1cm}w}^{(2)}\!-\!\fr{1}{6\pi })\!-\!2(\al^{-1}_3\!+\!\De_{3,\hs{0.1cm}w}^{(2)}\!-\!\fr{1}{4\pi })\!\r
\!+\!\fr{1}{6}\ln \ka \!-\!\fr{1}{3}\ln \fr{p}{\hat{p}}~.
\la{MX-SO10}
\eeq
We have taken GUT scale threshold corrections in one loop approximation.
The quantities $\De_{i,\hs{0.1cm}w}^{(2)}$ include weak scale  threshold corrections and 2-loop running effects for the gauge couplings, including Yukawa interactions.
Their values depend on the SUSY particle spectrum.
We carry out our analysis within the minimal $N=1$  SUGRA scenario  \cite{Chamseddine:1982jx} with family universal parameters.
While we vary these parameters to draw our conclusions, for concreteness, we consider the set of values:
$\{\tan \bt , m_0, m_{1/2}, \mu  \}=\{3, 1448.2~{\rm GeV}, 155.93~{\rm GeV}, 1~{\rm TeV} \}$, which
corresponds to $m_{\tilde{q}} \simeq 1.5$ TeV and  $m_{\tilde{W}} \simeq 130$ GeV . For these values we obtain
$\De_{i,\hs{0.1cm}w}^{(2)}=(0.6093, 0.4079, 1.167)$.
Using the third RGE equation for the $\alpha_i^{-1}$ we obtain for the present model $\alpha_G(M_X) \simeq \alpha_3(M_X) \simeq 1/20$.
The parameter $\ka \equiv M_8/M_3=2$ in our model (as opposed to $\ka =1$ in SUSY $SU(5)$).

It is important that the ratio $r\equiv M_{\Si }/M_X$ entering into Eq. (\ref{MX-SO10}) is constrained by symmetries of the model.
Using expressions for $M_{\Si }$, $M_A$ and $M_X$ presented earlier, we obtain:
\beq
r=\fr{M_{\Si }}{M_X}=\fr{4\lam_A'}{g^2}\fr{M_X}{M_*} \approx \l \fr{1}{15}-\fr{1}{300}\r ~.
\la{r-expr}
\eeq
The range for $r$ is obtained by noting that $\lam_A'$ is allowed by symmetries of the model and thus naturally expected to be of order one.
We have thus taken,  $1/5\stackrel{<}{_\sim }\lam_A'\stackrel{<}{_\sim }2$ (say), and have set $g^2\simeq g_{\rm GUT}^2\approx 0.63(1\pm 0.10)$,
$M_X\approx (0.6-1) \times 10^{16}$~GeV (see discussion after Eq. (\ref{p-pi-nu})), while $M_*\approx M_{\rm Pl}\simeq 2\tm 10^{18}$~GeV.
This restriction on $r$ will be an important ingredient in the derivation of an upper limit on $\Ga^{-1}(p\to e^+\pi^0)$.

Eliminating $p/\hat{p}$ from Eqs. (\ref{Meff}) and (\ref{MX-SO10})
we obtain a correlation between $M_{\rm eff}$ and $M_X$  for a given $r$:
\beq
M_{\rm eff}\simeq 10^{19}{\rm GeV}\hs{-0.05cm}\cdot \hs{-0.05cm}\l \fr{10^{16}{\rm GeV}}{M_X}\r^3\hs{-0.2cm}
 \l \fr{1/100}{r}\r \l \fr{3}{\tan \bt }\r \l \fr{0.6}{\eta_{\ga }}\r
\left \{\fr{\exp [2\pi (\De_{2,\hs{0.1cm}w}^{(2)}-\De_{3,\hs{0.1cm}w}^{(2)}-\de \al_3^{-1})]}{2.54\cdot 10^{-2}}\right \}\hs{0.05cm}~,
\la{corel-eq}
\eeq
where  $\eta_{\ga }\simeq 0.6$ accounts for the running of $\cos \ga $, and $\de \al_3^{-1}$ denotes the deviation of $\al_3^{-1}$
from its central value of $1/0.1176$. Note that the curly bracket on the right side of Eq. (\ref{corel-eq}) is fully determined for any given choice of the SUSY parameters and $\al_3(M_Z)$.  It turns out to be only mildly dependent
on variations of $m_0$ and $m_{1/2}$.
Since $M_{\rm eff}$  and $M_X$ respectively control
$d=5$ ($p\to \bar{\nu }K^+$) and $d=6$ ($p\to e^+\pi^0$) decay amplitudes, Eq. (\ref{corel-eq}) in turn provides a correlation between the rates of these two otherwise unrelated decay modes. Such a correlation exists in minimal SUSY $SU(5)$ as well \cite{Hisano:1994hb}, but that leads to predictions for $\al_3(M_Z)$ and $d=5$ proton decay rate which are inconsistent with experiments \cite{Murayama:2001ur}. Exceptions to this conclusion has been suggested in Ref.
\cite{Bajc:2002bv} which uses higher dimensional operators.  This however leads to large threshold corrections, making the apparent unification of gauge couplings with low energy SUSY somewhat coincidental. For review of proton decay in $SU(5)$ and in alternative scenarios see Ref. \cite{Nath:2006ut}.

\begin{figure}
\begin{center}
\leavevmode
\leavevmode
\vspace{-0.5cm}
\includegraphics{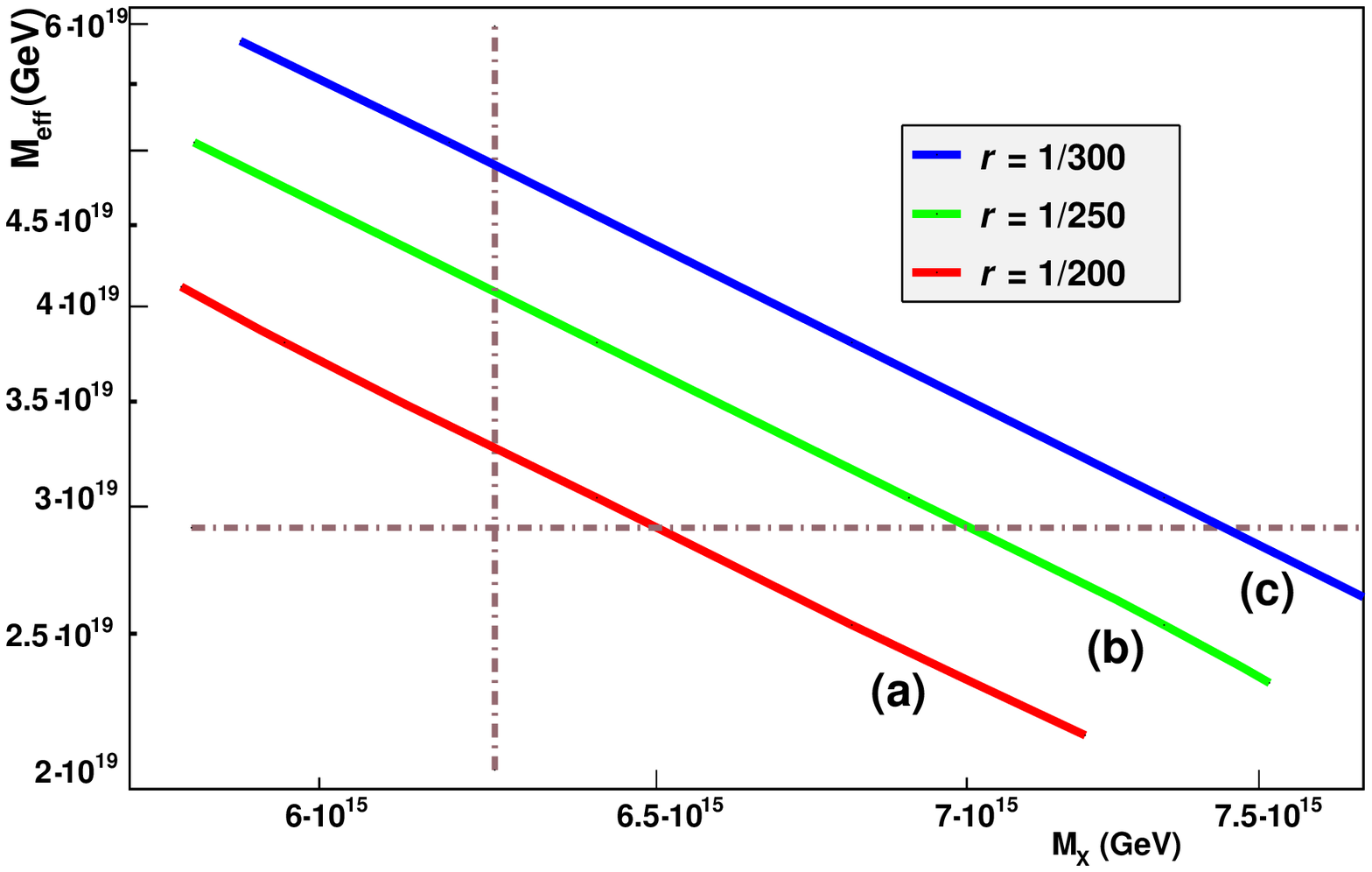}  
\end{center}
\vs{8.45cm}
\caption{Correlations between $M_{\rm eff}$ and $M_X$ for $\{\tan \bt , m_0, m_{1/2}, \mu  \}=\{3, 1448.2~{\rm GeV}, 155.93~{\rm GeV}, 1~{\rm TeV} \}$
(corresponding to $m_{\tilde{q}} = 1.5$ TeV, $m_{\tilde{W}} = 130$ GeV),
and $\al_3(M_Z)= 0.1176$.
{\bf (a)}: $r=1/200$. {\bf (b)}: $r=1/250$. {\bf (c)}: $r=1/300$. The vertical
and horizontal dashed lines correspond to the experimentally allowed lowest values of $M_X$ and $M_{\rm eff}$
which arise from limits on $\Ga^{-1}(p\to e^+\pi^0)$ and $\Ga^{-1}(p\to \bar{\nu }K^+)$ respectively, for central values of relevant parameters (see text).
}
\label{cor-case2Aal}
\end{figure}

Now, using expressions for proton decay rates (see below) one finds that the empirical lower limit on $\Ga^{-1}(p\to \bar{\nu }K^+)$ requires  that
$M_{\rm eff}\stackrel{>}{_\sim }2.91\cdot 10^{19}$~GeV
(for reasonable scenarios for the Yukawa couplings, see discussions in Sec. 4.1), while that on $\Ga^{-1}(p\to e^+\pi^0)$ requires (owing to Eq. (\ref{corel-eq}))
$r\stackrel{<}{_\sim }1/150$. Using the particular choice of SUSY parameters stated above,
and the ranges for $M_{\rm eff}$ and $r$ as given in Eqs. (\ref{Meff-range}) and (\ref{r-expr}),
we therefore illustrate our results on the correlation between
 $M_{\rm eff}$ and $M_X$  in Fig.  \ref{cor-case2Aal} by confining to the ranges $M_{\rm eff}\simeq (2.91-6)\tm 10^{19}$~GeV and  $r\simeq (1/200-1/300)$.\footnote{While such large values of $M_{\rm eff}$ lie in their natural ranges, they correspond via Eq. (\ref{Meff}) to rather small values of $\hat{p}/p\sim 10^{-4}$. Thus $\hat{p}/p\sim 10^{-4}$ is the only parameter of the model whose smallness remains unexplained on grounds of naturalness.}

Two crucial differences between our model and that of minimal SUSY $SU(5)$ (see Ref. \cite{Hisano:1994hb}--\cite{Nath:2006ut}) are:
(i) $M_{H_c}$ of $SU(5)$ is replaced by $M_{\rm eff}\cos \ga $ which can be significantly larger than $M_{\rm GUT}$ (for $\tan \bt \geq 3$ say); and
(ii) whereas the range for $r$ is severely restricted by symmetry considerations  for the $SO(10)$ model (see Eq. (\ref{r-expr})), this is not so for the minimal SUSY $SU(5)$ model.
These two distinctions make the $SO(10)$ model presented here more predictive for proton decay into $e^+\pi^0$ on the one hand, and viable on the other. In particular, as discussed below, with $\al_3(M_Z)$ being consistent with experiments, the $d=5$ proton decay rate is in full accord with the experimental limit.

\section{Nucleon decay}


There are two main mechanisms for proton decay corresponding to
$d=5$ and $d=6$, which respectively yield $p\to \bar{\nu }K^+$ and $p\to e^+\pi^0$ as the dominant decay modes.
Although apriori these two modes are largely independent, owing to the correlation given in Eq. (\ref{corel-eq}) and  Fig. \ref{cor-case2Aal},
they get linked in our model such that the observed lower limit on the inverse decay rate of either mode implies an  {\it upper limit}
on that of the other. The latter is found to be especially constrained for the $p\to e^{+}\pi^0$ mode.
The rates for $d=6$  decay modes $p\to e^{+}\pi^0$ and $p\to \bar{\nu }\pi^{+}$, which are largely independent of the details of Yukawa couplings and SUSY spectrum, are given by:
\beq
\Ga (p\!\to \!e^{+}\pi^0)\simeq \fr{m_p}{64\pi f_{\pi }^2}(1\!+\!D\!+\!F)^2\bar \al_H^2
\l \!\fr{g_X^2A_R}{M_X^2}\!\r^2\!\!f(p)~,~~
\Ga (p\!\to \!\bar{\nu }\pi^{+})\simeq 2\Ga (p\to e^{+}\pi^0)\fr{ f(p)\!-\!4}{f(p)}~,
\la{p-pi-nu}
\eeq
where $f(p)= 4+(1+1/(1+p^2))^2$.
Here $\al_H$  denotes the hadronic matrix element. Recent lattice calculation yields $\al_H\simeq 0.012~{\rm GeV}^3$ at $\mu =2$~GeV \cite{Aoki:2006ib}. $D$ and $F$ are chiral lagrangian
parameters with $D\simeq 0.8$, $F\simeq 0.47$. $g_X$ denotes the effective  $X, Y$ boson coupling at $M_X$.  The correlation curves (see Fig. \ref{cor-case2Aal}
for a representative case)
restrict $M_X$  in the range of about $(6-10)\tm 10^{15}$~GeV. Taking an average of the three gauge couplings, which nearly unify at $M_X$, lying  in the range  as given above
(see Fig. \ref{f:unif}), we obtain $\al_G(M_X)=g_X^2/(4\pi )\simeq (1/20) ( 1\pm 0.1)$, where the error reflects variations in the  GUT scale spectrum or equivalently in the parameters
of the superpotential lying in a natural range. The function $f(p)$ varies between the limits $8$ and $5$ as $p$ varies from $0$ to $\infty $; correspondingly
one obtains $\Ga (p\to e^{+}\pi^0)/\Ga (p\to \bar{\nu }\pi^{+})\simeq \l 1, 1.4, 2.5\r$ for
$\l p\stackrel{<}{_\sim }1/3~, p\approx 1~, p\gg 1\r $. The case of $p\to \infty $ represents the $SU(5)$ limit. If this branching ratio is measured to be significantly smaller than $2.5$, that would be strongly suggestive of $SO(10)$ (as opposed to $SU(5)$) grand unification.
The quantity $A_R$ in Eq. (\ref{p-pi-nu}) denotes the net renormalization of the $d=6$ operator, including short ($A_S^{d=6}\simeq 2.22$) and long distance effects ($R_L\simeq 1.25$).
In our model  $A_R\simeq 2.78$. From Eq. (\ref{p-pi-nu}) we get:
\beq
\Ga_{d=6}^{-1}(p\to e^{+}\pi^0)\simeq 1.0\times  10^{34}\hs{0.1cm}{\rm yrs}
\l \fr{0.012{\rm GeV}^3}{\al_H}\r^2 \hs{-0.15cm}\l \fr{2.78}{A_R}\r^2\hs{-0.15cm}\l \fr{5.12}{f(p)}\r \hs{-0.15cm}
\l \fr{1/20}{\al_G(M_X)}\r^2 \hs{-0.1cm}\l \fr{M_X}{6.24\times 10^{15}{\rm GeV}}\r^4.
\la{d6-p-to-epi}
\eeq
Allowing for uncertainty in $|\al_H|$ by $\pm 25\%$ (see discussion in Ref. \cite{Aoki:2006ib})
and that in $\al_G(M_X)$ by $\pm 10\%$, and letting $p$  vary in the theoretically favored range of $p\simeq 1$ to $p=10$,
we see that the empirical lower limit on $\Ga^{-1}(p\to e^{+}\pi^0)\stackrel{>}{_\sim }1.01\tm 10^{34}$~yrs \cite{Kobayashi:2005pe}
requires $M_X\stackrel{>}{_\sim }  \l 6.26\cdot 10^{15} {\rm GeV} \r \cdot \l 1\pm 0.14\r $,
where for the central value we have used $p=4$, and the errors are added in
quadratures.\footnote{This is only to indicate a reasonable range for
$(M_X)_{\rm min}$, which, however, is not used for our explicit predictions.}

Without further theoretical constraint on $M_X$, given that $\Ga_{d=6}^{-1}(p\to e^{+}\pi^0)\propto M_X^4$, there has been considerable uncertainty in the literature so far on $\Ga^{-1}(p\to e^{+}\pi^0)$, which has been quoted to lie in the range $\sim 10^{34}-10^{38}$~yrs \cite{pdg,Bajc:2002bv}. Such a range corresponds to (apriori reasonable) guesses on $M_X\sim M_{\rm GUT}\tm (1/3-3)\sim (0.7-6)\tm 10^{16}$~GeV, the higher values of $M_X$ being allowed by letting $r$
be arbitrarily small ($\stackrel{<}{_\sim }10^{-4}$ say). Lifetimes much exceeding $({\rm few}-10)\tm 10^{35}$~yrs would, however, be inaccessible
to next generation proton decay experiments. As discussed below, the correlation Eq. (\ref{corel-eq}) together with the restriction on $r$ (see Eq.
(\ref{r-expr})) would provide a much stronger constraint on $\Ga^{-1}(p\to e^{+}\pi^0)_{\rm max}$, which is fully accessible. To obtain an upper limit on $M_X$ and thus on
$\Ga^{-1}(p\to e^{+}\pi^0)$ we first need to discuss $d=5$ proton decay.

\subsection{Fermion masses and $p \rightarrow \overline{\nu}K^+$ decay rate}

In order to investigate $d=5$ proton decay, the Yukawa sector needs to be specified. We have found a self-consistent picture for fermion masses and mixings in the present setup in the same spirit as in Ref. \cite{Babu:2004dp,Babu:1998wi}.
Here, for the sake of completeness, we present only the gist of this
picture, we will return to a more complete presentation in \cite{inprep}.
We introduce a non-Abelian flavor symmetry $Q_4$ (the quaternionic group)
and assign the matter fields $16_{1,2}$ as a doublet of this group, $\orvec{16}=(16_1, 16_2)$. $16_3$ transforms trivially under $Q_4$. Two
$Q_4$ doublet flavon fields $\orvec{X}, \orvec{Y}$ both with VEVs along $(1, 0)$ direction are also utilized.
The $Q_4$ symmetry also enables us to successfully address the SUSY FCNC problem \cite{Pouliot:1993zm}.
With the ${\cal U}(1)_A$ charge assignments of $Q(\orvec{16})=-(1/2+3/k)$, $Q(\orvec{X})=3/k$ and $Q(\orvec{Y})=7/k$,
the relevant operators, in accord with the symmetry $SO(10)\tm {\cal U}(1)_A\tm Z_2\tm Q_4$, which generate effective Dirac Yukawa couplings, are:
$16_3 16_3H, \fr{\orvec{X}}{M_*}\orvec{16} 16_3H,
\fr{SZ^2A}{M_*^4} \orvec{16} \orvec{16}H,\,
\fr{Z^3C}{M_*^4}\orvec{16} \orvec{16}C'$,
$\fr{AC\orvec{Y}}{M_*\lan Z\ran^2} \l \orvec{16}\cdot 16_3+\right. $ $\left. 16_3\cdot \orvec{16}\r C'$, and
$\fr{AC}{M_*^2\lan Z\ran^2}(\orvec{X}\orvec{16})(\orvec{Y}\orvec{16})C'$.
The higher order operators, suppressed by powers of $1/M_*$, and in the last two cases by $1/\lan Z\ran^2$ as well, may be generated by quantum gravity and in part by exchange of additional heavy vector-like states ( see \cite{inprep} for details).
The resulting mass matrices for the quarks and charged leptons at the GUT scale have the form:
\beq
\begin{array}{ccc}
 & {\begin{array}{ccc}
\hs{-0.8cm} u^c_1 & \hs{0.1cm} u^c_2 & \hs{0.1cm} u^c_3
\end{array}}\\ \vspace{1mm}
M_u= \hs{0cm}
\begin{array}{c}
 u_1 \\  u_2 \\  u_3
 \end{array}\!\!\!\!\!\hs{-0.1cm}&{\left(\begin{array}{ccc}

 0  & \ep' & 0
\\
-\ep' &0 & \si
 \\
 0 & \si  & 1
\end{array}\right)}m_U^0 ~,
\end{array}
\begin{array}{ccc}
 & {\begin{array}{ccc}
\hs{-0.6cm} d^c_1(e_1) & \hs{0.8cm} d^c_2(e_2) & \hs{0.8cm} d^c_3(e_3)
\end{array}}\\ \vspace{1mm}
M_{d(e)}= \hs{-0.1cm}
\begin{array}{c}
 d_1(e_1^c) \\  d_2(e_2^c) \\ d_3(e_3^c)
 \end{array}\!\!\!\!\!\hs{-0.1cm}&{\left(\begin{array}{ccc}

 0 & \ka_{d(e)}\ep' +\eta'& 0
\\
-\ka_{d(e)}\ep' -\eta' &\ka_{d(e)}\xi_{22}^d & \si +\ka_{d(e)}\ep
 \\
 0 & \si +\ka_{d(e)}\ov{\ep }  & 1
\end{array}\right)}m_D^0~,
\end{array}
\la{mass}
\eeq
where $m_D^0=m_U^0 \cos \ga /\tan \bt$, $\ka_{d}=1$ and $\ka_e=3$.  Eqs. (\ref{mass}) provide
a constrained system with fewer parameters than observables.  A consistent
fit for
all masses and mixing parameters as well as observed CP violation is obtained with the choice
$\si =0.0508, \ep =-0.0188+0.0333i, \ov{\ep }=0.106+0.0754i ,
\ep'=1.56\cdot 10^{-4}, \eta'=-0.00474+0.00177i,
\xi_{22}^d=0.014 e^{4.1i}$ at the GUT scale. Following  renormalization in going down to low
energies (with $m_t(m_t) = 160$ GeV and $\tan\beta = 3$), these values reproduce the
central values of the charged lepton masses.  In addition, for the quark masses we obtain
\begin{eqnarray}
m_u(2~{\rm GeV})&=&3.55~{\rm MeV},~m_c(m_c)=1.15~{\rm GeV}, \nonumber \\
m_d(2~{\rm GeV})&=&6.45~{\rm MeV},~m_s(2~{\rm GeV})=137.6 {\rm MeV},~m_b(m_b) = 4.67~{\rm GeV}.
\end{eqnarray}
For the CKM mixings we obtain at $\mu=M_Z$,
\begin{equation}
|V_{us}|=0.225~,~~|V_{cb}|=0.0414~,~~|V_{ub}|=0.0034~,~~|V_{td}|=0.00878,~~
\ov{\eta }=0.334~,~~\ov{\rho }=0.12~.
\la{CKMfinal}
\end{equation}
Thereby we get $\sin 2\bt =0.663$. All these are in a good agreement with experiments.

Let us now briefly discuss the neutrino
sector.
The relevant operators, responsible for generating heavy Majorana masses for the right-handed neutrinos are:
$\fr{Z^{k-4}\orvec{Y}^2}{M_*^{k-3}M_{N''}^2}\orvec{16}^2\bar C^2$, $\fr{Z^{k-2}}{M_*^{k-2}M_{N'}^2}\orvec{Y}\cdot \orvec{16}16_3\bar C^2$, and
$\fr{Z^{k-1}S}{M_*^{k-1}M_N^2}16_3^2\bar C^2$.
Here, $M_N$, $M_{N'}$ and $M_{N''}$ represent masses of additional singlets which turn out to lie in the range of
(few - $100$)$M_{\rm GUT}$ \cite{inprep}.
We assume that in the first of these
couplings the $Q_4$ contraction $\orvec{Y}^2\orvec{16}^2$ is in the $1'$ channel. The heavy Majorana mass
matrix $M_R$ is given by:
\beq
%
\begin{array}{ccc}
 & {\begin{array}{ccc}
\hs{-0.7cm} \nu^c_1 & \hs{-0.1cm} \nu^c_2 & \hs{-0.1cm} \nu^c_3
\end{array}}\\ \vspace{1mm}
M_R= \hs{-0.1cm}
\begin{array}{c}
  \nu^c_1 \\  \nu^c_2 \\ \nu^c_3
 \end{array}\!\!\!\!\!\hs{-0.1cm}&{\left(\begin{array}{ccc}

 b & 0& 0
\\
0 &b & a
 \\
 0 & a & 1
\end{array}\right)}M_0~.
\end{array}
\la{Dir-Maj}
\ee
The Dirac mass matrix $M_{\nu D}$ at GUT scale can be obtained from $M_u$ (see Eq. (\ref{mass})) by the replacement $\ep'\to -3\ep'$.
We can take the two dimensionless parameters $(a,\,b)$ and the mass parameter $M_0$
as input to fix $\sq{\De m_{\rm atm}^2}$, $\te_{12}$ and $\te_{23}$. Two observables, viz.,
$\sq{\De m_{\rm sol}^2/\De m_{\rm atm}^2}\simeq m_2/m_3$ and $\te_{13}$, will then be predictions
of the model. The structures given in Eq. (\ref{Dir-Maj}) are valid at GUT scale.
Applying renormalization, including threshold effects due to the different $\nu^c$ masses,
with $\te_{12}\simeq 30^o$ and $\te_{23}\simeq 43^o$ as inputs, we obtain $m_2/m_3\simeq 0.13$
and $\te_{13}\simeq 3.6^o$ as predictions.
Such a fit is realized by choosing $a=0.0252e^{-0.018i}$, $b=1.61\cdot 10^{-6}e^{-1.592i}$,
and $M_0=1.89\cdot 10^{13}$~GeV. The corresponding $\nu^c$ masses are $(M_{R1}, M_{R2}, M_{R3})=(3.04\tm 10^7, 1.2\tm 10^{10}, 1.79\tm 10^{13})$~GeV.
These results include all the relevant RG running effects.
One sees broad, although not precise, agreement with
data.  We consider this fit, which provides large neutrino
oscillation angles, together with small quark
mixing angles as well as observed CP violation
as fairly successful and highly nontrivial, especially
in a quark-lepton unified framework with a
stabilized doublet-triplet splitting.

With the Yukawa couplings specified,
the inverse of the sum of partial  decay widths, in $p\to \bar{\nu}K^+$, is computed to be:
$$
\Ga^{-1}_{d=5}(p\to \bar{\nu}K^+)=3.5\tm 10^{33}{\rm yrs}
\l \fr{0.012{\rm GeV}^3}{|\bt_H|}\r^2\l \fr{6.91}{\bar{A}_S^{\al }}\r^2
\l \fr{1.25}{R_L}\r^2\l \fr{M_{\rm eff}}{3.38\tm 10^{19}{\rm GeV}}\r^2\tm
$$
\beq
\tm \l \fr{m_{\tl{q}}}{1.5{\rm TeV}}\r^4\l \fr{130}{m_{\tl{W}}}\r^2
\l \fr{3.1}{K_{d=5}^{\nu }}\r~.
\la{d5-nuK}
\eeq
Here $K_{d=5}^{\nu }$ denotes a sum of contributions to the total decay rate from
the three neutrino flavors, reflecting the dependence of the $d=5$ operator
on the Yukawa couplings:
 $K_{d=5}^{\nu }\equiv|A_{\nu_e}|^2+|A_{\nu_{\mu }}|^2+|A_{\nu_{\tau }}|^2$.
Each individual $A_{\nu_i}$ receives contributions from three types of diagrams leading to the $d=5$ operator: (a) those with only the first two families in the external legs, (b) those having the quark doublet of the third family
in just one external line, and (c) those having the same as in (b) in two external lines.
The last two contributions incorporate the short distance renormalization of the $d=5$ operator that arises through the running of the top quark Yukawa coupling, from the GUT scale to the weak scale. Contributions from all three diagrams are found to be important, especially for $|A_{\nu_{\mu }}|$
and $|A_{\nu_{\tau }}|$. The net result is that  $|A_{\nu_e}|\sim {\cal O}(10^{-1})$,  $|A_{\nu_{\mu }}|\sim |A_{\nu_{\tau }}|\sim {\cal O}(1)$ and
$K_{d=5}^{\nu } \simeq 3.1$ \cite{inprep}.
$\bar{A}_S^{\al}$ in Eq. (\ref{d5-nuK}) denotes the short distance RGE factor for the $d=5$ operator, corresponding to the running from the GUT scale to the weak scale, that arises purely from the gauge interactions, without the effects of the top quark Yukawa coupling. Note that $\bar{A}_S^{\al}$ defined here differs from the RGE factor $A_S$ defined conventionally  \cite{Hisano:1994hb}
in that $A_S$ includes the effect of the running of $m_cm_s$ in going from low energies to the GUT scale, $\bar{A}_S^{\al}$ does not.Thus,
$A_S=\bar{A}_S^{\al}J$, where $J=(m_cm_s)_{\rm GUT}/(m_cm_s)_{\mu }\sim
{\cal O}(10^{-1})$ for $\mu =2$~GeV (with low $\tan \bt \sim 3$ to $10$).

We can now discuss the derivation of an upper limit on $M_X$ and thus on $\Ga^{-1}(p\to e^{+}\pi^0)$. Owing to Eq. (\ref{corel-eq}) this would correspond
to the minimum allowed value of $M_{\rm eff}$. Now, taking conservatively $m_{\tl{q}}\stackrel{<}{_\sim }1.5$~TeV and the experimental lower limit
$m_{\tl{W}}\stackrel{>}{_\sim }125$~GeV, the observed lower limit on $\Ga^{-1}(p\to \bar{\nu }K^+)\stackrel{>}{_\sim }2.8\tm 10^{33}$~yrs
\cite{Kobayashi:2005pe} yields (via Eq. (\ref{d5-nuK})): $(M_{\rm eff})_{\rm min}\stackrel{>}{_\sim }2.91\tm 10^{19}$~GeV. This in turn yields by using Eq.
(\ref{corel-eq}) (with $|\beta_H| = 0.012$ GeV$^3$, the lowest value of $r=1/300$ and $\tan \bt =3$): $(M_X)_{\rm max}\stackrel{<}{_\sim }(5.16, 7.02, 9.45)\tm 10^{15}$~GeV for
$\al_3(M_Z)=(0.1156, 0.1176, 0.1196)$.  Thus, if we use central values of $\alpha_3(M_Z), \alpha_H, \beta_H$
and $\alpha_G(M_X)$ with $p=4$, the upper limit on $\Gamma^{-1}(p \rightarrow e^+ \pi^0)$ (using Eq.
(\ref{p-pi-nu})) would be $1.61\times 10^{34}$ yrs.  If the uncertainties in all these parameters are stretched to
their extremes, {\it each in a direction so as to prolong the lifetime},  the stated upper limit
could increase by atmost a factor of $10.8$.  Considering that all the uncertainties having such extreme values, and
in the same direction, to be very unlikely, we would regard something like the geometric mean of the
two upper limits corresponding to the central and extreme values of the parameters to be a more realistic,
yet conservative, upper limit for the lifetime.  We thus predict:
\vs{-0.2cm}
\beq
\Ga_{d=6}^{-1}(p\to e^{+}\pi^0)\stackrel{<}{_\sim }5.3\times 10^{34}~{\rm yrs}~.
\la{d6-limit}
\eeq
If $m_{\tl{q}}<1.5$~TeV, or $m_{\tl{W}}>125$~GeV, or $r>1/300$, or $\tan \bt >3$, the upper limit would of course decrease further significantly.\footnote{While $\Gamma^{-1}(p \rightarrow e^+ \pi^0)$ given by
Eq. (\ref{p-pi-nu}) does not explicitly depend on $m_{\tilde{q}}$, $m_{\tilde{W}}$, $r$ and $\tan\beta$, the
{\it upper limit} on $M_X$ and thereby $\Gamma^{-1}(p \rightarrow e^+ \pi^0)$ does depend on these parameters via
the correlation Eq. (\ref{corel-eq}).  The latter relates $(M_X)_{\rm max}$ to $(M_{\rm eff})_{\rm min}$ and thereby
to the empirical lower limit on $\Gamma^{-1}(p \rightarrow \overline{\nu} K^+)$ which depends on $m_{\tilde{q}}$ and $m_{\tilde{W}}$.
Because of this, the upper limit given in Eq. (\ref{d6-limit}) should in fact be multiplied by an approximate factor
$(m_{\tilde{q}}/1.5~{\rm TeV})^{8/3} ~(125 ~{\rm GeV}/m_{\tilde{W}})^{4/3}~ [(1/300)/r]^{4/3}~(3/\tan\beta)^{4/3}$.}
Thus, the upper limit shown above on $\Ga^{-1}(p\to e^{+}\pi^0)$, stemming from Eq. (\ref{corel-eq}), is a robust and novel feature of the model.
The predicted lifetime is accessible to proposed megaton size water Cherenkov (or equivalent) detectors.

Reversing the procedure given above, we can derive an upper limit on $M_{\rm eff}$ and thereby on
$\Gamma^{-1}(p \rightarrow \overline{\nu} K^+)$.  Owing to Eq. (\ref{corel-eq}), this would correspond
to the minimum allowed value of $M_X$ and $r$.  Using central values of $|\alpha_H|$ and
$\alpha_G(M_X)$ with $p \approx 4$ (for concreteness), the observed lower limit on
$\Gamma^{-1}(p \rightarrow e^+ \pi^0) \geq 1.01\times 10^{34}$ yrs \cite {Kobayashi:2005pe} yields
via Eq. (\ref{d6-p-to-epi}): $(M_X)_{\rm min} \geq 6.26 \times 10^{15}$ GeV.  This in turn yields by using Eq.
(\ref{corel-eq}) (with $m_{\tilde{q}} = 1.5$ TeV, $m_{\tilde{W}} = 130$ GeV and the lowest values of $r \approx 1/300$):
($M_{\rm eff})_{\rm max} \leq (1.627,~4.105, ~10.02) \times 10^{19}~{\rm GeV} (3/\tan\beta)$ for
$\alpha_3(M_Z) = (0.1156,~0.1176,~0.1196)$.\footnote{Note that these values of $(M_{\rm eff})_{\rm max}$ are quite consistent with the estimate given in
Eq. (\ref{Meff-range}).}  If we use central
values of the parameters and the spectrum as noted above, with $p \approx 4$ and $\tan\beta \geq 3$, the upper
limit on $\Gamma^{-1}(p \rightarrow \overline{\nu} K^+)$ (using Eq. (\ref{d5-nuK})) would be $5.16\times 10^{33}$ yrs.
Allowing for uncertainties in the parameters in a combined manner, analogous to the case of $d=6$ lifetime, we thus
obtain\footnote{We have checked \cite{inprep} that the rate for $d=5$ proton decay generated by
Planck scale operators are sufficiently suppressed, owing to symmetries present in the model,
so as not to disturb the upper limit quoted in Eq. (\ref{nubarKplus}).}
\begin{equation}\label{nubarKplus}
\Gamma^{-1}(p \rightarrow \overline{\nu} K^+) \stackrel{<}{_\sim } (3.1\times 10^{34}~{\rm yrs})\times
\left({m_{\tilde{q}} \over 1.5~{\rm TeV}}\right)^4\left({130~{\rm GeV} \over m_{\tilde{W}}}\right)^
2~(3/\tan\beta)^2~.
\end{equation}
In Eq. (\ref{nubarKplus}) the mild dependence of the curly bracket of Eq. (\ref{corel-eq}) on $m_{\tilde{q}}$ and $m_{\tilde{W}}$ is not exhibited.
The actual lifetime is  likely to be significantly lower than few $\times 10^{34}$ yrs if $M_{\rm eff}$
is not stretched to its upper limit (corresponding to e.g., $M_X > (M_X)_{\rm min}$, or $r > 1/300$, or
$\tan\beta > 3$ and/or $\alpha_3(m_Z) < 0.1196$), or if $m_{\tilde{q}} < 1.5$ TeV, or $m_{\tilde{W}} > 130$ GeV.
We thus find that not only the $p\to e^{+}\pi^0$ mode, but very
likely even the
$p\to \bar{\nu }K^+$ mode should be observable by improving the current experimental sensitivity by about a factor of five to ten.

Some important details concerning the present work, including those pertaining to the issues of fermion masses and mixings,
and some variants as regards the cancelation of the ${\cal U}(1)_A$ Fayet-Iliopoulos term,
 will be presented in a forthcoming longer paper \cite{inprep}.

In summary, we have presented a class of supersymmetric $SO(10)$ models with low dimensional Higgs system that fully resolves all the naturalness issues of doublet-triplet splitting, including stability against higher order operators, generation of $\mu $-term of order $m_{susy}$, and proton stability. The threshold corrections in these models are found to depend only on a few effective parameters, making the scenario very predictive.
An intriguing feature of these models is the correlation equation and the corresponding
constrained upper limit on $\Ga^{-1}(p\to e^{+}\pi^0)$.
We find that in this class of models proton decays into both $e^{+}\pi^0$ and very likely  $\bar{\nu }K^+$
as well should show with an improvement in the current
sensitivity by about a factor of five to ten. The building of a megaton water Cherenkov detector (or equivalent) would thus be most welcome.

\vs{0.1cm}

We would like to thank Takaaki Kajita and Edward Kearns for helpful
communications.
K.S.B. and Z.T. are
supported in part by US Department of Energy, Grant Numbers DE-FG02-04ER41306 and DE-FG02-ER46140.
J.C.P. is supported in part by the US Department of Energy, contract Number DE-AC02-76SF00515 and by DOE grant No. DE-FG02-96ER41015.
Z.T. is also partially supported by GNSF grant 07\_462\_4-270.

\vs{-0.6cm}

\bibliographystyle{unsrt}


\end{document}